# Explainable AI and Machine Learning Towards Human Gait Deterioration Analysis

**Abdullah Alharthi, King Abdullah University of Science and Technology (KAUST), Thuwal, Saudi Arabia. abdullah.alharthi@kaust.edu.sa**

**Abstract**

Gait analysis, an expanding research area, employs non-invasive sensors and machine learning techniques for a range of applications. In this study, we concentrate on gait analysis for detecting cognitive decline in Parkinson's disease (PD) and under dual-task conditions. Using convolutional neural networks (CNNs) and explainable machine learning, we objectively analyze gait data and associate findings with clinically relevant biomarkers. This is accomplished by connecting machine learning outputs to decisions based on human visual observations or derived quantitative gait parameters, which are tested and routinely implemented in current healthcare practice. Our analysis of gait deterioration due to cognitive decline in PD enables robust results using the proposed methods for assessing PD severity from ground reaction force (GRF) data. We achieved classification accuracies of 98% F1 scores for each PhysioNet.org dataset and 95.5% F1 scores for the combined PhysioNet dataset. By linking clinically observable features to the model outputs, we demonstrate the impact of PD severity on gait. Furthermore, we explore the significance of cognitive load in healthy gait analysis, resulting in robust classification accuracies of 100% F1 scores for subject identity verification. We also identify weaker features crucial for model predictions using Layer-Wise Relevance Propagation. A notable finding of this study reveals that cognitive deterioration's effect on gait influences body balance and foot landing/lifting dynamics in both classification cases: cognitive load in healthy gait and cognitive decline in PD gait.

## 1. Introduction

Human gait is the unique manner in which each individual walks. Gait involves a cyclical sequence of movements of both lower limbs that can be described as a series of transitions between states [1]. There is significant interest in sensing and recognizing human gait for various applications, including healthcare [2], [3], sports [4], [5], biometrics [6],[7],[8], and human-robot interaction [9], [10]. Gait provides important information about a person's physical and physiological characteristics, such as weight, gender, health, and age.

The first focus of this work was on gait in Parkinson's disease (PD) patients. PD is a common neurodegenerative disorder caused by loss of neurons in the midbrain [11]. While a definitive PD diagnosis requires pathological examination, limited reliable neuropathological criteria prevent conclusive diagnosis during life. Historically, only 80-90% of clinical PD diagnoses have been confirmed at autopsy [12]. PD is associated with reduced life expectancy, partly due to the 25-40% of patients who develop dementia [11], [12]. Currently, PD diagnosis and severity rating rely primarily on clinical evaluation and subjective surveys using scales like the Unified Parkinson's Disease Rating Scale, Hoehn and Yahr staging, and Schwab and England activities of daily living [13]. Gait deviation is a hallmark of PD, and disease progression increases fall risk [11]. However, in early PD subtle gait changes may lead to inconclusive visual evaluation, partly because slow walking and short strides can also indicate age, mood, or other conditions [12]. Consistent with this, PD manifests as tremor, rigidity, and bradykinesia. Gait analysis aids PD diagnosis and tracking, though current methods are semi-subjective.

The second focus was on gait under cognitive load ("dual tasks"). Gait patterns vary within and between individuals due to factors like dual-tasking, environment, energy optimization, and emotional state. Dual tasking alters gait in all individuals, indicating higher-level cognitive input is required for gait [15], [16]. Humans continuously adjust their gait to minimize energy cost, even for small savings [17], [18],[19]. Emotional states like happiness or fear also impact gait [20]. However, gait inconsistency could enable biometric verification if signatures under cognitive load prove individually distinct. Gait analysis may detect and characterize age-related cognitive decline using inexpensive sensors, aiding diagnosis of mobility impairment, increased fall risk [21], [22], and disorders like Alzheimer's disease or vascular dementia [23].

The motivation here is to categorize cognitive load decline in PD and identify subjects using gait under cognitive load. However, gait is a sequence of periodic events that naturally vary slightly in all individuals due to temporary psychological or lasting physiological conditions. Therefore, visual observation, harmonic analysis, Fourier decomposition, or their combination may not adequately represent gait cycle nonlinearity and nonstationary. Recent advances in deep learning offer an alternative for processing ground reaction force (GRF) signals to achieve reliable gait classification. Deep artificial neural networks (ANNs) using raw sensor data implement automatic feature extraction, avoiding subjective feature engineering [24]. This enables analyzing gait deterioration, though a known ANN limitation is opacity, hindering understanding [25] of predictions in terms of domain knowledge [26]. This limits feedback to improve sensor design and data processing. To address this, layer-wise relevance propagation (LRP) [26] relates ANN predictions to input data. By conservation of relevance [27], [26], it produces relevance maps attributing portions of predictions to raw input, identifying important areas. LRP shows success in image classification [28], [29] and gait-based subject identification [30].

This work presents a deep convolutional neural network (DCNN) to analyze GRF data and categorize PD and dual-task gait deterioration. We validate classification using LRP relevance scores to add noise. Sensor fusion uses CNNs to classify gait, and explainable CNNs [26] relate results to observable gait events, identifying the most relevant for each class. This uses the defined cyclic patterns in healthy gait to query what parts are essential for recognition and which act as background (irrelevant) in CNN processing. LRP interprets CNN predictions and identifies highest-weighted gait events for recognition.

## 2. Background

To enhance the understanding of our work, we provide an overview of the key concepts involved in analyzing gait. This includes a concise summary of the theoretical foundations of popular machine learning frameworks and the essential procedures for training, validating, and testing convolutional neural networks. Additionally, we introduce the Layer-Wise Relevance Propagation (LRP) approach, which improves the interpretability and explainability of artificial neural networks. Moreover, we provide a brief overview of the latest research on gait parameters and an update on the literature since our last review paper on this topic [ 24].

### 2.1 Deep Convolutional Neural Networks (CNNs)

CNNs are state-of-the-art machine learning models that have proven effective in a variety of classification tasks, providing valuable insights into complex data. These networks can learn high levels of abstraction and features from large datasets by applying convolutional operations to the input data. CNNs are composed of convolution layers, perceptron layers, pooling layers, and normalization layers. A set of filters and weights are shared among these layers. The convolutional layers output a feature map that is automatically extracted from the raw input data, followed by a perceptron layer based on neurons that map the features to an output. Each convolutional layer is then followed by pooling layers that reduce computational cost by decreasing the size of representation and making the convolution layer output more robust. The convolutional neural network shown in Figure 1.

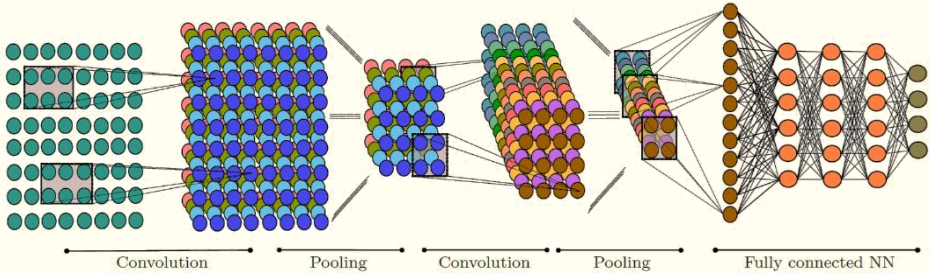

Figure 1. Illustration of a convolutional neural network with two-dimensional convolutional blocks and subsampling as pooling operation followed by Perceptron layer.

### 2.2 Convolutional Layer

Convolution operation is performed on the input data and a filter or kernel to produce a feature map as shown in figure 1. In this process, the filter slides over input data and performs convolution. The convolution operations output feature maps. The learning occurs at the kernel by updating the kernel values during training. At the end, the feature map output consists of different feature maps produced by different kernels as convolution layer output. An activation function is utilized to produce nonlinear feature maps that can be optimized during training to pass valuable neurons values to the next layer. A mathematical representation of a convolution operation in one dimension with an input vector $x$, a kernel $\omega$ with $i, d$ to denote iterators, and $(\circ)$ to denote the element-wise multiplication, can be expressed as $C(i)$ with $i$ is the index of an element in the new feature map (ch. 9 [31]):

$$C(i) = (\omega \circ x)[i] = \sum_d x(i-d)\, \omega(d) \text{-} \qquad (1)$$

Gait is captured as two dimensions signal as spatial and temporal, therefore the convolution operation in eq. 1 can be extended to two dimensions. Such that the spatiotemporal input is a large set of data points, and the kernel is a set of data smaller in size than the input. Then the convolution operation slides the kernel over the input and computes element-wise multiplication and add the values in a smaller future map. With a 2-D input $x$ and a 2-D kernel $\omega$ with $(i,j)$, $(d,k)$ are iterators, the mathematical representation of a convolution in two dimensions can expressed as $C(i,j)$ with $(i,j)$ is the index of an element in the new feature map [31]:

$$C(i,j) = (\omega \circ x)[i,j] = \sum_d \sum_k x(i-d, j-k)\omega(d,k) \qquad (2)$$

## 2.3 Backpropagation

It is short for "backward propagation of errors", it is an algorithm based on gradient descent. The method moves in a reverse order from the output layer to the input layer while calculating the gradient of the error function based on the network weights, the aim is to minimize $J(\theta)$ using an optimal set of parameters in $\theta$. It is based on performing the partial derivative to minimize the cost function. The partial derivative is expressed as $\frac{\partial}{\partial \theta_{i,j}^l} J(\theta)$. The output layer calculates the error of the network layers $L$ with: $ð^{(L)} = \alpha^{(l)} - y$, such that the error of node $j$ in layer $l$ is denoted as $ð_j^{(l)}$ and the activation of node $j$ of layer $l$ is denoted as $\alpha_j^{(l)}$ and $y$ is the output of the output layer, then the backpropagation can be expressed for neural networks as:

$$ð^{(L)} = ((\theta^{(l)})^{(T)} ð^{(l+1)}) \circ \alpha^{(l)} \circ (1 - \alpha^{(l)}) \qquad (3)$$

Here the ð values of the output layer L are calculated by multiplying the ð values in the next layer (in reverse direction) with the θ matrix of layer l, hence T denote matrix. We then perform elementwise multiply ($\circ$) with the $g'$ which is the derivative of the activation function, which is evaluated with the input values given by $z^{(l)}$. Where $g'(z^{(l)}) = \alpha^{(l)} \circ (1 - \alpha^{(l)})$.

The partial derivatives needed for backpropagation is performed by multiplying the activation values and the error values for each training example t and m is the number of training data as:

$$\frac{\partial}{\partial \theta_{i,j}^l} J(\theta) = \frac{1}{m} \left[ \sum_{t=1}^m \alpha_j^{(t)(l)} ð_j^{(t)(l+1)} \right] \qquad (4)$$

## 2.4 Evaluation Measure

The widely used accuracy measure for gait analysis is the confusion matrix [32]. It is a table to visualize the number of predictions classified correctly and wrongly for each class. The table consists of true positive, true negative, false positive, and false negative classification occurrences. One of the advantages of the confusion matrix display is that it is straightforward to identify the decision confusions, thus possibly concluding on the quality of the data involved. It shows each class prediction as follows:

**True positive,** TP: It is the number of positive classes correctly predicted as positive.

**True negative,** TN: It is the number of negative classes correctly predicted as negative.

**False positive,** FP: It is the number of negative classes incorrectly predicted as positive.

**False negative,** FN: It is the number of positive classes incorrectly predicted as negative.

From this confusion matrix table, the number of predictions classified correctly and wrongly are used to calculate different rates of measure to evaluate the performance of a machine learning model. Performance measures, such as accuracy, recall, precision and F1 values of a model can be defined using the following equations.

**Accuracy**: indicator of the ratio between the correctly predicted data to total number of samples in the dataset, defined as: $\frac{TP+TN}{TP+TN+FP+FN}$

**Recall**: the proportion of positive classes identified correctly, defined as: $\frac{TP}{TP+FN}$

**Precision**: the fraction of positive cases correctly identified over all the positive cases predicted, defined as: $\frac{TP}{TP+FP}$

**F1 Score**: the harmonic mean of Precision and Recall, defined as: $\frac{2*Precision*Recall}{Precision+Recall}$

There are popular evaluation measures used for classification problems such as Area Under the Curve (AUC) and Receiver Operating Characteristic (ROC). In this thesis we use the confusion matrix over the Area Under the Curve (AUC) because the number of TP, TN, FP and FN samples are values of interest to understand the confusion in gait classes for further analysis using LRP.

## 2.5   Layer-Wise Relevance Propagation (LRP)

LRP [25], [26] [27] is a backward propagation method which identifies which parts of the ANN input vector carry most weight in the model prediction. In this thesis we quantify the contribution of a single component of an input $x_i$ (in our case, a sensor signal at a specific time frame) to the prediction of $f_c(x)$ ($c$ denote a class of gait) made by the DCNN classifier $f$. The outputted gait class prediction is redistributed to each intermediate node via backpropagation until the input layer. The LRP outputs a "heat map" over the original signal to highlight the signal sections with the highest contributions to the model prediction, e.g., the data sections with the maximum variability given the classes. We first note that a neural network consists of multiple layers of neurons (feature maps in the case of a convolution layer), where neurons are activated as follows [26]:

$$a_k = \sigma(\sum_j a_j \omega_{jk} + b_k) \qquad (5)$$

Here, $a_k$ the neuron activation and $a_j$ is the activation of the neuron in the previous layer in forward direction; $\omega_{jk}$ denote the weight received in forward direction by neuron $k$ from neuron $j$ in the previous layer and $b_k$ is the bias. The sum is computed over all the $j$th neurons that are connected to $k$th neuron. $\sigma$ is a nonlinear monotonically increasing activation function. These activations, weights, and biases are learned by the DCNN during supervisory training. During training, the output $f_c(x)$ is evaluated in a forward pass and the parameters ($\omega_{jk} + b_k$) are updated by back-propagating using model error. For the latter, we base our computations on categorical cross entropy [33].

The LRP approach decomposes the DCNN output for a given prediction function of gait class $c$ as $f_c$ for input $x_i$ and generates a "relevance score" $R$ for the $i$th neuron received from $R_j$ for the $j$th neuron in the previous layer which is received from $R_k$, for the $k$th neuron in the lower layer (see figure 2), where the relevance conservation principle is satisfied as:

$$\sum_i R_{i \leftarrow j} = \sum_j R_{j \leftarrow k} = \sum_k R_k = f_c(x) \qquad (6)$$

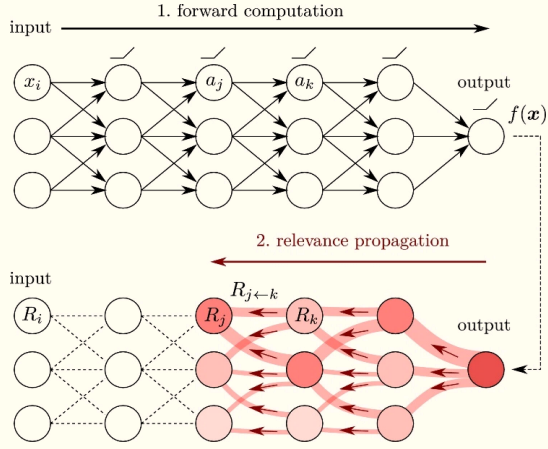

Figure 2. DNN and LRP signal processing flow. Red arrows indicate the relevance propagation flow [26].

The LRP starts at the DCNN output layer after removing the *Softmax* layer. In this process, a gait class $c$ is selected as an input to LRP, and the other classes are eliminated. The backpropagation for unpooling for the pooling layer is computed by redirecting the signal to the neuron for which the activation was computed in the forward pass. As a generalization, consider a single output neuron $i$ in one of the model layers, which receives a relevance score $R_j$ from an lower layer neuron $j$, or the output of the model (class $c$). The scores are redistributed between the connected neurons throughout the network layers, based on the contribution of the input signals $x_i$ using the activation function (computed in the forward pass and updated by back-propagating during training) of neuron $j$ as shown in figure 2. The latter will hold a certain relevance score based on its activation function and passes its value to consecutive neurons in the reverse direction. Finally, the method outputs relevance scores for each sensor signal at a specific time frame. These scores represent a heat map, where the high relevance scores at specific time frames highlight the areas that contributed the most to the model classifications. There are other propagation rules such as (αβ-rule) [26].

$$R_j = \sum_k \left( \alpha \, \frac{a_j \omega_{jk}^+}{\sum_j a_j \omega_{jk}^+} - \beta \, \frac{a_j \omega_{jk}^-}{\sum_j a_j \omega_{jk}^-} \right) R_k \qquad (7)$$

Where each sum corresponds to $R_{j \leftarrow k}$ a relevance message and $a_j \boldsymbol{\omega}_{jk}^+$ and $a_j \boldsymbol{\omega}_{jk}^-$ denote the positive and negative part of $a_j \, \omega_{jk}$ respectively. The parameters α and β are chosen so that α−β =1 and β ≥ 0. A propagation rule can be chosen by selecting β = 0 to in result the following rule :

$$R_j = \sum_k \frac{a_j \omega_{jk}^+}{\sum_j a_j \omega_{jk}^+} R_k \qquad (8)$$

There are other stabilizing terms that can be used to avoid divisions by zero as explained in [26],[27]. For the LRP-$\gamma$ rule, let the neurons interconnection be as follow [30]:

$$a_k = \max\left(0, \sum_{0,j} a_j \, \omega_{jk}\right) \qquad (9)$$

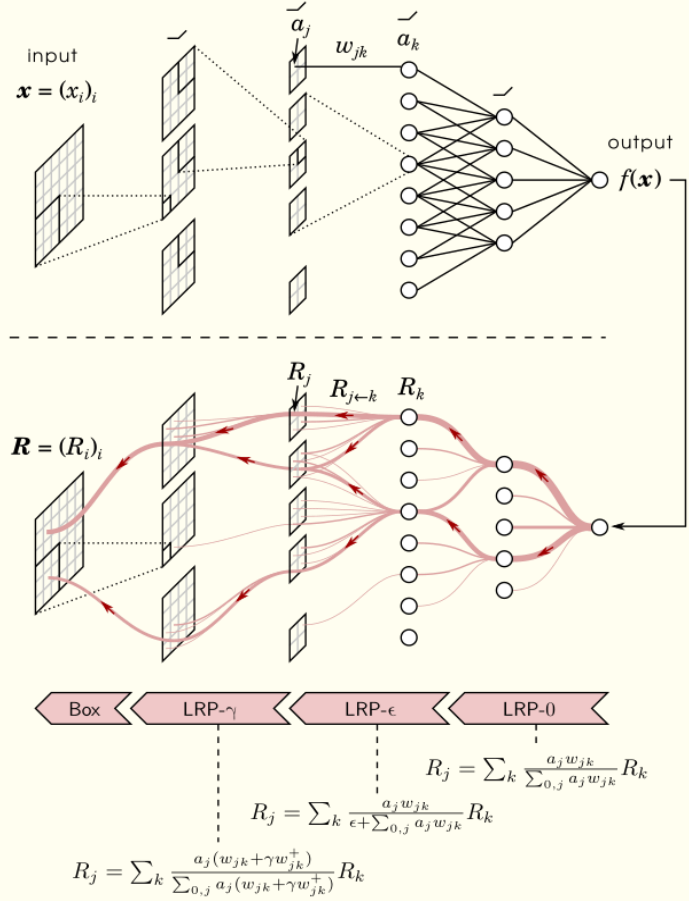

Figure 3. Illustration of the LRP propagation procedure applied to a neural network. The prediction at the output is propagated backward in the network, using various propagation rules, until the input features are reached. The propagation flow is shown in re[30].

Here, $a_j$ denote input activation and $\omega_{jk}$ denote the weight received by neuron $j$ from neuron $k$ in the above layer. The sum is computed over all neurons $j$ in the lower layer plus a bias term $\omega_{0k}$ with $a_0 = 1$. The LRP-$\gamma$ rule as shown in figure 3 is given by:

$$R_j = \sum_k \frac{a_j(\omega_{jk}+\gamma\omega_{jk}^+)}{\sum_{0,j} a_j(\omega_{jk}+\gamma\omega_{jk}^+)} R_k \tag{10}$$

## 2.6 Gait Parameters

Gait can be perceived as a transformation of a brain activity to muscle contraction patterns resulting in a walking sequence. It is a chain of commands generated in the brain and transmitted through the spinal cord to activate the lower neural center, which will consequently result in muscle contraction patterns assisted by sensory feedback from joints, muscles and other receptors to control the movements. This will result in the feet recurrently contacting the ground surface to move the trunk and lower limbs in a coordinated way, delivering a change in the body center-of-mass position.

Gait is a sequence of periodic events characterized as repetitive cycles for each foot [27]. Each cycle is divided into two phases (see figure 4):

a) **Stance Phase** (approximately 60% of the gait cycle, with the foot in contact with the ground). This phase is subdivided into four intervals (**A, B, C, D**).

b) **Swing Phase** (approximately 40% of the gait cycle with the foot swinging and not in contact with the ground). This phase is subdivided into three intervals (**E, F, G**).

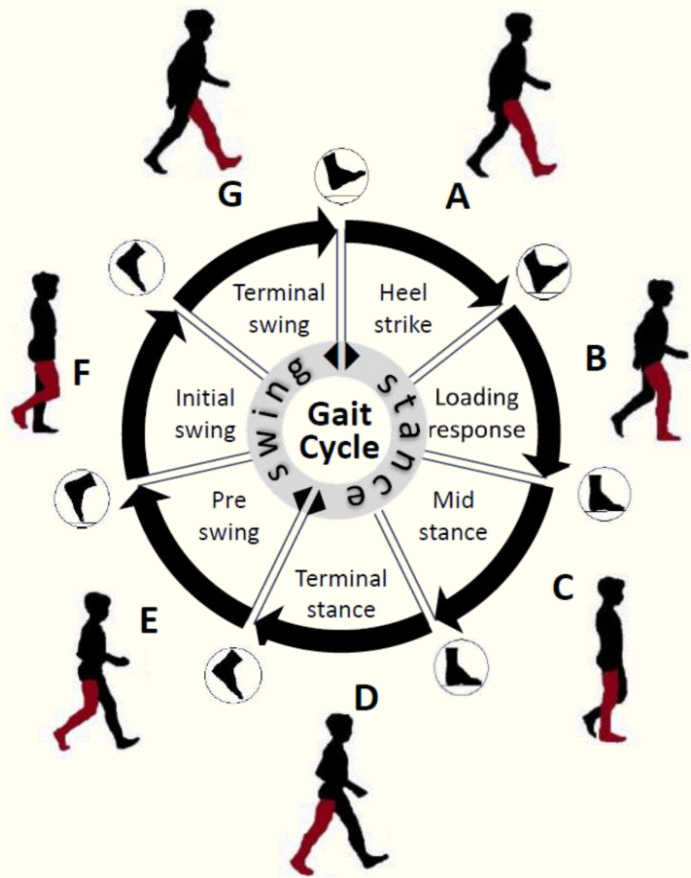

Figure 4. Important gait events and intervals in a normal gait cycle.

**Stance Phase:**

A. Heel strike or Initial contact: It starts the moment the foot touches the ground, and it is the initial double-limb support interval. In the case of the right foot leading, the double support starts with left foot being on the ground when the right foot heel makes initial contact and finishes when the left foot leaves the ground with the left toe-off prepared to swing. At the end of this interval, the body weight is completely shifted onto the stance (leading) limb. This term is adopted in clinical psychology to denote the contact of the heel of the extended limb with the walking surface.

B. Loading response or Foot flat: This is a single support interval following the initial double support interval. The bodyweight is transferred on to the supporting limb. The trunk is at its lowest position, the knee is flexed, and a plantarflexion occurs at the ankle.

C. Mid-stance: This is a single support interval between opposite toe-off and heel-off. It starts from elevation of opposite limb until both ankles are aligned in coronal plane. The trunk is in its highest point and slowing its forward speed. The body center-of-mass is aligned with the forefoot (ball of the foot).

**D.** Terminal stance or Heel-off: It begins when the supporting heel rises from the ground in preparation for opposite swing. The trunk is sinking from its highest point, the knee has extant peak near the time of heel rise and ankle has dorsiflexion after heel rise.

**Swing Phase:**

**E.** Pre-swing: This is the second double-limb support interval. The opposite initial contact occurs, and the hip is beginning to flex, the knee is flexing, and the ankle is at plantarflexion. The toe is in last contact before the swing, finishing the push-off started in interval **D**.
**F.** Initial swing and Mid-swing: This interval begins with the toe-off into single support and starting to swing. The body weight is shifted to the opposite forefoot. In this instant, the knee joint gets the maximum flexion. The hip is flexing and the limb advances in preparation for a stride.
**G.** Terminal swing: This is the last interval of gait cycle and the end of the swing phase. The interval begins at maximum knee flexion and ends with maximum extension of the swinging limb forward. The hip continues flexion and the knee extends in regard to gravity, the ankle continues dorsiflexion to end neutral, ready for the heel strike.

With regard to the above gait events, the following parameters of human gait are usually analyzed in clinical settings [34] for healthcare tasks, using various sensing and data processing methods:

- Cadence or rhythm (number of steps per unit time)
- Stride length
- Velocity
- Direction of leg segments
- Step angle
- Swing time for each foot
- Step width
- Support time
- Ground Reaction Force (GRF)
- Electrical activity produced by muscles
- Momentum and forces
- Body posture

## 2.7 Gait Related Work

Gait recognition constitutes a significant research field with a steady progression in the variety of studies. It mainly focuses on biometric and gait impairments for possible diagnoses and monitoring the severity of injuries or diseases. Further to our recent topical review [24] on gait recognition, gait recognition literature in the past two years has been focused on solving the view- and clothing invariant problems for video sequences with more advanced machine learning methods, such as generative adversarial network [34],[35]. Zhang et al. [34] designed a view transformation generative adversarial network (VT-GAN) to translate gaits between any two views only using single model. The proposed VT-GAN model includes three models namely, a generator, a discriminator, and similarity preserver. The model achieved a competitive result on the cross-view gait task using CASIA-B dataset described in [24]. The other use of GAN in the work of Babaee et al. [35] focuses on gait recognition from incomplete gait cycle. GAN is used in this work to reconstruct complete GEIs from incomplete GEIs. The GAN architecture is composed of a generator with an autoencoder to reconstruct a complete GEIs from incomplete GEIs, the other component is a discriminator. The model is evaluated on OU-ISIR dataset [24] and achieved a competitive result in the case of incomplete gait cycle. A more recent use of GAN is in the work of Chen et al. [36], the proposed Multi-View Gait GAN (MvGGAN) for cross-view gait. The models are evaluated on CASIA-B [24] and OUMVLP [37] datasets and achieving improved results for gait recognition performance in reality scenes.

Recent work on wearable and floor sensors has been applied for medical applications such as the impact of muscle fatigue on gait characteristics [38], health monitoring [39] and age-related differences [40]. In recent work on wearable sensors by Turner et al. [41], LSTM network is proposed to process and classify pressure sensors signals. The sensors were placed inside the shoes and participants were asked to walk eight walking trails. The aim of this study was to analyze artificially induced gait alterations. The results are promising for a potential use in the diagnosis of gait abnormalities or other neuromuscular movement disorders in patients. In a recent work on wearable sensors. Tran et al. [42], proposed multi-model LSTM and CNN to classify IMUs spatiotemporal signals. The proposed models outperformed previous results on the whuGAIT [43] and OU-ISIR [44] datasets using hybrid network.

Table 1. Datasets subject's discerption.

| Subjects | Number | Male | Female | Group |
|---|---|---|---|---|
| PD patients | 29 | 20 | 9 | Ga [45] |
| Healthy Subjects | 18 | 10 | 8 | Ga [45] |
| PD patients | 29 | 16 | 13 | Ju [46] |
| Healthy Subjects | 26 | 12 | 14 | Ju [46] |
| PD patients | 35 | 22 | 13 | Si [47] |
| Healthy Subjects | 29 | 18 | 18 | Si [47] |

Table 2 Number of subjects with the severity rating.

| Severity (0) Healthy | Severity (2) | Severity (2.5) | Severity (3) | Group |
|---|---|---|---|---|
| 18 | 15 | 8 | 6 | Ga[45] |
| 26 | 12 | 13 | 4 | Ju [46] |
| 29 | 29 | 6 | 0 | Si [47] |

## 3. Material and Methodology

### 3.1 Parkinson's disease data

To assess GRF data from PD patients we used the open access benchmark from PhysioNet.org [48]. It consists of data from 93 PD patients (mean age: 66.3 years; 63% men), as detailed in table 1, with different levels of PD progression, as detailed in table 2. Data from 73 healthy controls (mean age: 66.3 years; 55% men) are also present. The dataset consists of GRF measurements collected as participants walked for approximately two minutes. Each subject had eight sensors placed underneath each foot to measure force [N] as a function of time. The output of the 16 sensors was recorded at 100 frames per second. Also, the sum of the eight sensors of each foot is added to each subject sample and the timestamp, yielding 19 columns in total. The data set was collected by three research groups, namely: Ga group [45], Ju group [46] and Si group [47] with the sub-parts of the dataset named after these groups. The Ju and Si groups recorded usual healthy walking at a self-selected speed. The Ga repeated this, and included additional samples for each subject, where they performed dual task while walking [45].

### 3.2 Cognitive Load Data

The iMAGiMAT footstep imaging system is an original Photonic Guided-Path Tomography floor sensor head [49],[50],[51],[52]). It can record unobtrusively temporal samples from a number of strategically placed distributed POF sensors on top of a deformable underlay of a commercial retail floor carpet. Each sensor comprises of low cost POF (step-index PMMA core with fluorinated polymer cladding and polyethylene jacket, total diameter 1mm, NA=0.46) terminated with a LED (Multicomp OVL-3328 625nm) at one end and a photodiode (Vishay TEFD4300) at the other. The sensors constitute a carefully designed set to allow collaborative sensor fusion and deliver spatiotemporal sampling adequate for discerning gait events. The 1m x 2m area system is managed by 116 POF sensors, arranged in three parallel plies, sandwiched between the carpet top pile and the carpet underlay: a lengthwise ply with 22 POF sensors at 0º angle to the walking direction and two independent plies, each consisting of 47 POF sensors, arranged diagonally at 60º and -60º respectively (see [49], figure 6 therein). The electronics is contained in a closed hard-shell periphery at carpet surface level and is organised in 8-channel modules: LED Driver boards as well as input transimpedance amplifier boards to receive the data and send it to a CPLD (Complex programmable logic device) to reformat the data for processing by a Raspberry pi single board computer for export via Ethernet/WiFi. The operational principle of the system is based on recording the deformation caused by the GRF variations, as bending affects the POF sensors transmitted light intensity is affected by surface bending. This captures the specifics of foot contact and generates robust data without constraints of speed or positioning anywhere on the active surface.

Twenty-one physically active subjects aged 20 to 40 years, 17 male and 4 females, without gait pathology or cognitive impairment, participated in this experiment. The study was carried out under the University of Manchester Research Ethics Committee (MUREC), ethical approval number 2018-4881-6782. All participants were informed about the data recording protocol in accordance with the ethics board general guidelines and each subject written consent was obtained prior to experiments. Each participant was asked to walk normally, or while performing cognitively demanding tasks, along the 2 m length direction of the iMAGiMAT sensor head. The captured gait data is unaffected by start and stop, as it is padded on both ends with unrecorded several gait cycles before the first footfall on the sensor. With a capture rate of 20 timeframes/s (each timeframe comprising the readings of all 116 sensors), experiments yielded 5s long adjacent time sequences, each containing 100 frames. The recorded gait spatiotemporal signals were able to capture around 4 to 5 uninterrupted footsteps at each pass.

Five manners of walking were defined as normal gait plus four different dual tasks, and experiments were recorded for each subject, with 10 gait trials for each manner of walking in a single assessment session; thus the total number of samples is $10 \times 5 = 50$ per-subject. The five manners of walking are defined as follows:

- Manner 1, Normal Gait: walking at normal self-selected speed.

- Manner 2, Gait while listing to a story: audio input through headphones, then answer questions after gait recording is completed.

- Manner 3, Gait with serial 7 subtractions: normal walking speed attempted, while simultaneously performing serial 7 subtractions (count backward in sevens from a given random 3-digit number).

- Manner 4, Gait while texting: normal walking speed attempted, while simultaneously typing text on a mobile device keyboard.

- Manner 5, Gait while talking: walking at normal self-selected speed while talking or answering questions.

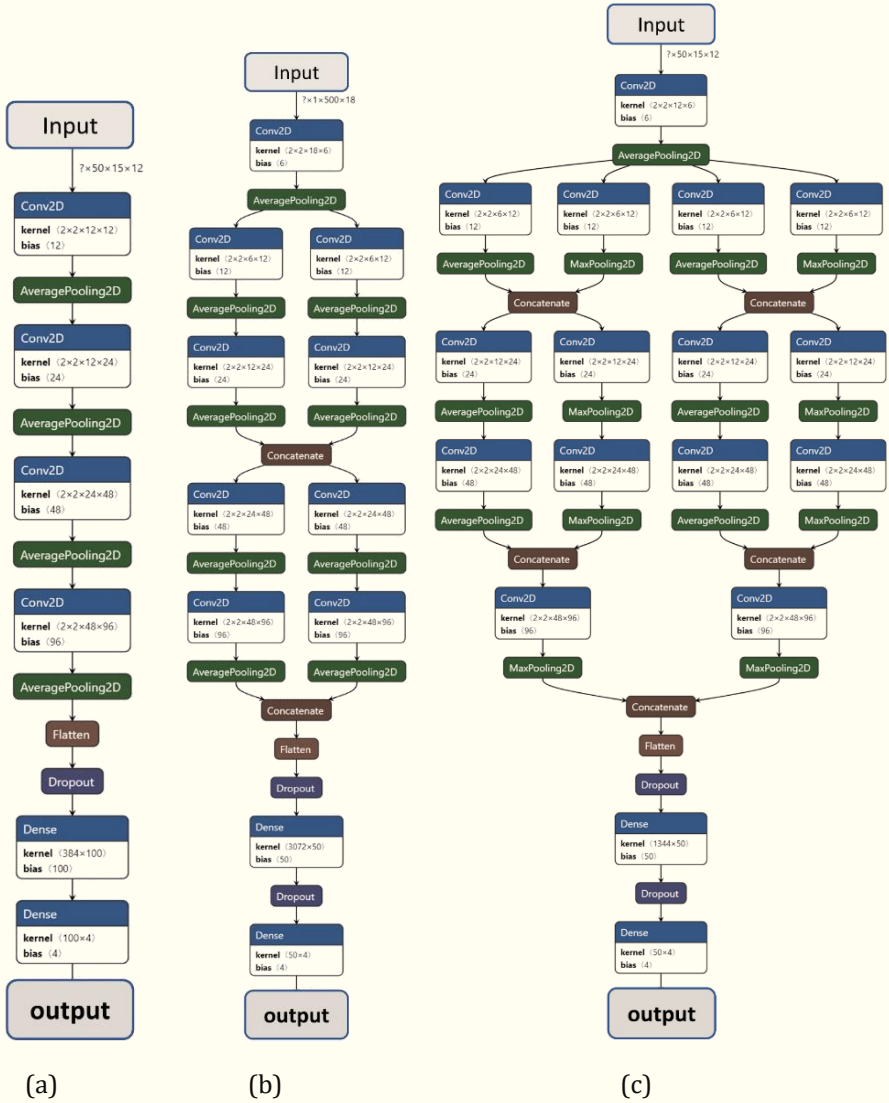

Figure 5. Proposed DCNN architectures; a) Single DCNN, b) Parallel DCNN, c) Quadruplets DCNN. The color-coding of boxes: convolution layers and fully connected layer (blue); pooling layers (green); concatenation layers and flattening layers (brown); dropout layers (navy); input at the top and softmax output layers (gray). The diagrams are generated using Neutron from GitHub repository based on the models' weights and biases.

### 3.3 Perturbation

Human gait is inconsistent between individuals, and even for a single individual, requiring models to be reliable and robust to variance in the input data. Applied to the LRP analysis, the interpretation of important input data points needs to be robust to noise and variance in the input data stream. Considering this, a random perturbation noise analysis on the LRP relevance score can aid in the choice of LRP method, as well as in designing a DCNN model resilient to noise due to the inconsistency of gait. This perturbation analysis is achieved as described in the following sections.

The "greedy" iterative procedure proposed in [53] allows to choose the appropriate LRP method and to evaluate the quality of gait classification relevance scores. This is achieved by step-wise removal of information from the spatiotemporal input signal. At each step, regions with the highest relevance scores are replaced by Gaussian noise by a "most relevant first" (MoRF) approach [53]). The change in the model performance is then monitored at each step by running the model to re-predict the test data with the perturbation accumulated by that step. The most desirable LRP method is the one with the strongest drop in accuracy at the first few steps [187], where the most relevant information is removed by the perturbations, and slower decline further as less important regions are removed. The accuracy drop is quantified using the area over the most relevant perturbation curve (AOPC) [53].

In terms of assessing the significance of DCNN model architecture, this task is achieved by progressively removing the highest relevance scores yielded by the best LRP method selected using the method described above and testing the model performance by re-predicting on the test data for each model. Models which substantially drop performance after only a few perturbation steps are considered to be the most amenable to exploiting LRP. This is because the decline in performance allows to assert that those few removed regions are critical for accurate classification performance, therefore it is indicative of meaningful relationships between input patterns and learnt classes. In contrast, if removing a region with little impact on the classification performance, the implication is that it is of lesser interest in terms of seeking such relationships.

### 3.4    Proposed DCNN Architectures

The classification of gait ground reaction force (GRF) signals is a challenging task that requires the use of advanced machine learning techniques. In previous work, the authors of this study experimented with several deep convolutional neural network (DCNN) models to process and classify spatiotemporal 3D matrices of raw sensor signals. The researchers' extensive experimentation led them to identify three different network architectures that showed promising results. The architectures are shown in Figure 5 and are detailed in the following sections.

#### 3.4.1    DCNN

A 2D-DCNN model (figure 5(a)) built for PD severity classification consists of four convolutional layers, each followed by an average pooling and two fully connected layers, yielding a total of 10 stacked layers. The four convolutional layers have *n* channels each that assign one frame of the input to a single convolutional layer channel. The convolutional layers use a stride of 1, same-padding, and a 2 x 2 filter for the average-pooling layers.

#### 3.4.2    Parallel DCNN

This network architecture (figure 5(b)) has been proposed specifically in our previous work [150] to process GRF signals. It is inspired by inception neural network architectures [154], the aim is to have filters with learnable parameters operate on the same level to recognize the salient parts in the sample. The network consists of two stages with parallel streams fused with concatenation layers, where each stream has its weights and biases launched uniformly and updated during training via backpropagation. The network is topped up with fully connected layers and a softmax layer, giving a total of 18 stacked layers.

#### 3.4.3    Quadruplet DCNN

The quadruplet network shown in figure 5 (c) is an original model and its architecture implemented with multi-parallel streams by generalizing Siamese [54] and triplet networks [55]. This network consists of convolutional layers, max-pooling and average-pooling with each stream has its activations, weights, and biases launched uniformly and updated separately via backpropagation. The goal of this network is to learn the spatial sensor signals and temporal signals separately and simultaneously (similar to inception neural network [56]), using two types of pooling layers. This allows the capturing of gait pattern and the ability of the network to generalize on unseen data.

## 4. DCNN Implementation and Results

All algorithms for LRP computation are implemented in Python 3.7.3 programming language using Keras 2.2.4, TensorFlow 1.14.0 and iNNvestigate GitHub repository [57]. All codes are run using a desktop with intel core i7 6700 CPU @3.4 GHz. After data standardization, the deep CNN model is applied on the dataset in order to test the validity of the algorithms for identifying gait signatures. We compared the CNN predictions to manually labelled ground truth in several experiments, PD severity staging, individuals' identity and the changes to normal gait incurred by cognitive load. The models' classification performance is evaluated using confusion matrices. The performance of the LRP methods is examined in detail in the discussion subsection.

### 4.1 Experiment (1) on PD Gait Data

### 4.1.1 Data Pre-Processing

Each sample recorded in the dataset contains 19 columns of data with varying column length, as for some subject's gait was recorded for a longer time (12119 frames) than for others (less than 1000 frames). In order to make the input data length consistent, the datasets were split to equal size parts of 500 frames such that single long recording are divided to several chunks of 500 frames. The timestamp columns were deleted as it doesn't report information about gait. The final sample size is 18 columns and 500 rows or frames, as shown in figure 6. This choice is justified since the gait cycle is approximately one second and the sample captures heel strike and toe off for both feet over five gait cycles. The input dataset is a tensor with dimensions $m \times 500 \times 18$ where m= 2698 for Ga group, 2198 for Ju and 1509 Si group (see example sequences in figure 6). The input is reshaped for the 2D-DCNN as $K \times 50 \times 15 \times 12$ building upon our previous work with different inputs and algorithms [58].

Data standardization is performed as a pre-processing step to reduce the redundancy and dependency among the data, such that the estimated activations, weights, and biases will update similarly rather than at different rates during the training process. The standardization involves rescaling the distribution of values with mean at zero and rescaling the standard deviation to unity.

$$\widehat{x_{n,s}} = \frac{x_{n,s} - \mu(x_{n,s})}{\vartheta(x_{n,s})} \qquad (11)$$

Here $\widehat{x_{n,s}}$ is PD data rescaled such that $\mu$ is the mean values and $\vartheta$ is the standard deviation. Next, the dataset is randomly split into training 60%, hold-out validation 20% and testing 20% with a *random state* parameter with different seed.

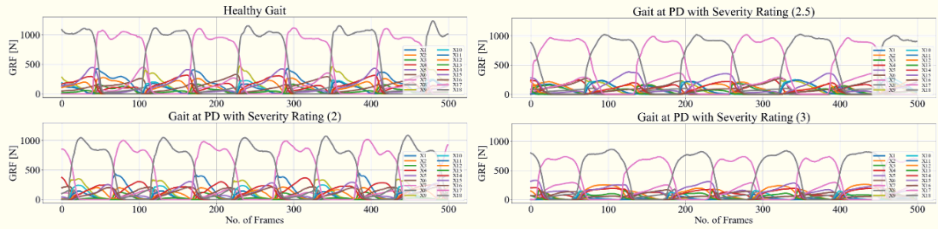

Figure 6. Example GRF data recorded at 100 frames per second for healthy subject and subjects with PD severity ratings 2, 2.5, and 3, with sample length of 500 timeframes $\widehat{x_{n,s}} = [x_1 \; \cdots \; x_{18}] \in \mathbb{R}^{500\times18}$. The signals with lower amplitude ($\widehat{x_{n,s}} = [x_1 \; \cdots \; x_{16}]$<500 s) represent pressure sensor signal under each foot (different colors for each of the 8 sensors). In each sample, the calculated sum of the 8 sensor outputs for each foot is also shown additionally ($x_{17}$&$x_{18}$>500 N, different colors for left and right foot).

Table 3. Models F1-Score for each Dataset and F1-Score, Mean and Standard Error with Datasets Combine.

| CNN Model | Ga | Ju | Si | A Seed 42 | A Seed 100 | A Seed 200 | A Seed 2020 | A $m$ | A $St$ |
|---|---|---|---|---|---|---|---|---|---|
| Single | 98% | 98% | 98% | 95% | 93% | 96% | 96% | 95% | 0.70% |
| Parallel | 96% | 97% | 96% | 96% | 95% | 95% | 96% | 95% | 0.28% |
| Quadruplet | 97% | 97% | 98% | 95% | 94% | 94% | 95% | 94% | 0.28% |

A: Ga U Ju U Si; *ST*: Standard error; *m*: Mean performance

### 4.1.2  Feature Learning and Classifications

The three models are trained, validated, and tested separately, using a batch size of 200 samples for each iteration, 200 epochs are optimal to train the models, determined by a trial way (on the three datasets combined) and the error using categorical cross-entropy. A method for stochastic optimization (ADAM [59]) is used to train the proposed models. The optimizer parameters are adjusted as follows: $\alpha$ =0.002, $\beta1$ =0.9, $\beta2$ =0.999, $\varepsilon$=1e-08. Where $\alpha$ is the learning rate or the proportion that weights are updated; $\beta1$ is the exponential decay rate for the first moment estimates; $\beta2$ is the exponential decay for the second-moment estimates; $\varepsilon$ is a small number to avoid any division by zero in the implementation.

The loss computed by the categorical cross-entropy in every iteration is used to validate the models and update the weights and biases. To improve the model's performance a regularization method is utilized together with dropouts, as shown in figure 5. The models are trained, validated and tested, three times separately (for each dataset) and four times (datasets combined) with different random state, to test the models' ability in classifying the three datasets and with the later combined. Accuracy is reported as confusion matrices in figure 7, as well as precision, recall and f1-score. The mean performance and standard errors of the different random state runs are reported in table 3.

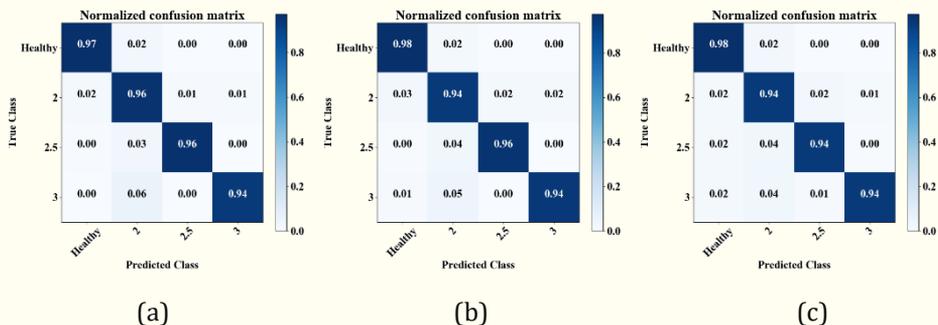

(a) (b) (c)

Figure 7. Models' predictions on 1281 sample are shown as confusion matrices: a) Single DCNN, b) Parallel DCNN, c) Quadruplet DCNN.

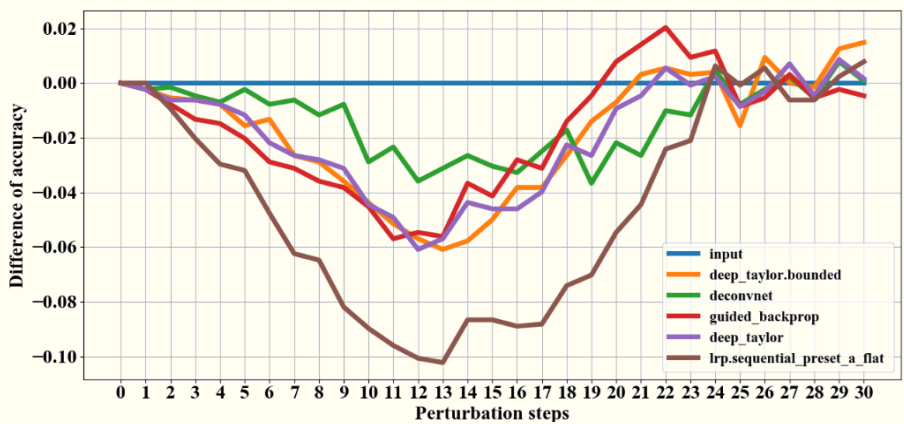

Figure 8. LRP method selection by perturbation steps progressively removing information with the highest relevance scores. Steeper initial decrease indicates better identification of gait events with most weight in the classifications.

### 4.1.3 LRP and Model selection

A number of LRP methods were tested with the three DCNNs models. All the models returned the same results; therefore, we report the result from single DCNN, to identify the best performing backpropagation method implemented in iNNvestigate GitHub repository: Deep Taylor [27], Deep Taylor bounded [60], deconvnet (deconvolution) [61], guided backprop (guided backpropagation) [62], and LRP sequential preset a flat (LRP-SPF) [60]. The DCNN classification accuracy is evaluated for each of the above LRP methods separately, by performing a sequence of perturbation steps, as described in section 3.3. (Progressively replacing MoRF regions of size 7×7, representing 0.544% (0.00544 = 7×7/ 50×15×12) of the input stream, with Gaussian noise), and observing for each LRP method the cumulative change in the model performance. The baseline for comparison is established by replacing regions of the input data with random Gaussian noise regions instead of replacing the regions based on LRP methods. Next, we subtract the LRP maps accuracy from the input with the randomly replaced regions accuracy to show only the LRP accuracy change. As shown in figure 8, in our case the LRP curves recover after around the 15th perturbation step, because the remaining spatiotemporal regions are less and less relevant and the baseline accuracy is reached around the 25th perturbation step, as then all remaining regions are unimportant for the classification. As expected, the exhibited rate of change is proportional to the importance of the information perturbed at each step.

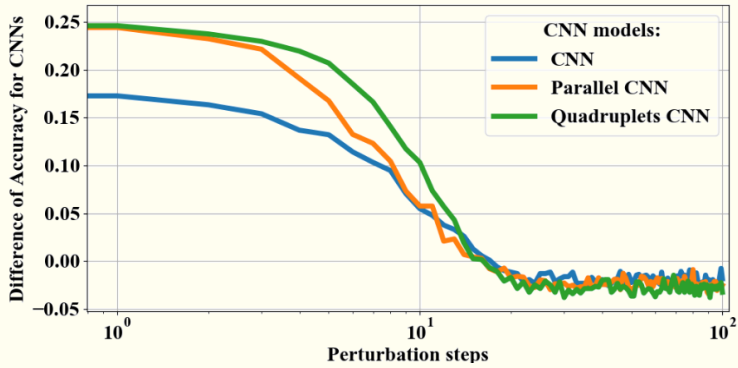

Figure 9. Semi-log plot of the perturbation effect on the proposed DCNNs architectures. The decline in accuracy results from progressively removing information from the input data based on LRP-SPF and re-predict, at each step, 100 steps total.

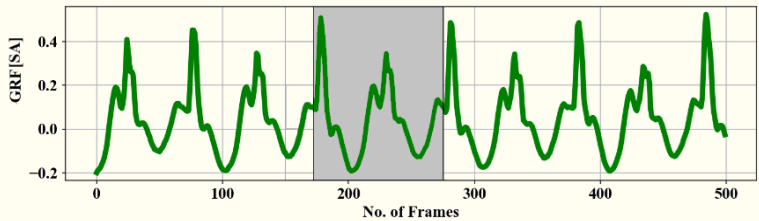

(a)

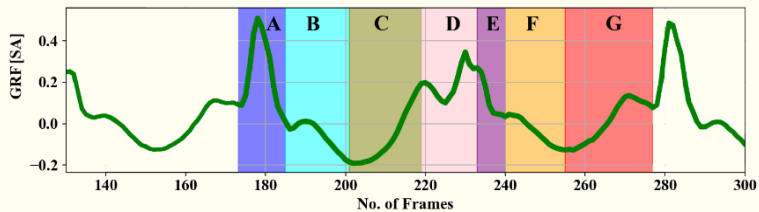

(b)

Figure 10. Gait events processed SA (see equation 13) signal top. The highlighted gray area in (a) is explained in (b) based on gait events for one foot from figure 4 as: A- Heel strike, B- Loading response or flat foot, C- Mid-stance or single support, D- Terminal stance or heel rising, E- Pre-swing or double-limb support, F- Initial swing and Mid-swing or toe-off, G-Terminal swing.

The choice of a DCNN model most suitable for LRP is justified by utilizing the same MoRF protocol [53], whereby each of the three DCNN models is perturbed step wise with Gaussian noise by progressively replacing MoRF regions of size 7×7 and re-predict gait class for 100 steps. In contrast with the LRP method selection illustrated in figure 8, instead of comparing with a baseline, here the rate of decline in accuracy (by removing the mean to show only the rate of change) with subsequent perturbation steps is used to identify the model with the steepest drop based on each model returned 100 classification accuracies, as manifested in figure 9. This prediction drop takes place at a faster rate in models where the classification uses data patterns in more compact regions within the gait cycle sequence. This allows more straightforward identification of the gait cycle events corresponding to such regions, with reference to the standard cycle presented in figure 4. Among the three proposed models, the Parallel DCNN (see figure 7.b) experiences the steepest decrease in accuracy with perturbation, as figure 9 shows, therefore is the preferred candidate to attempt the identification of gait events most vulnerable to gait deterioration due to PD. Table 3 summarizes the performance of the three DCNN models.

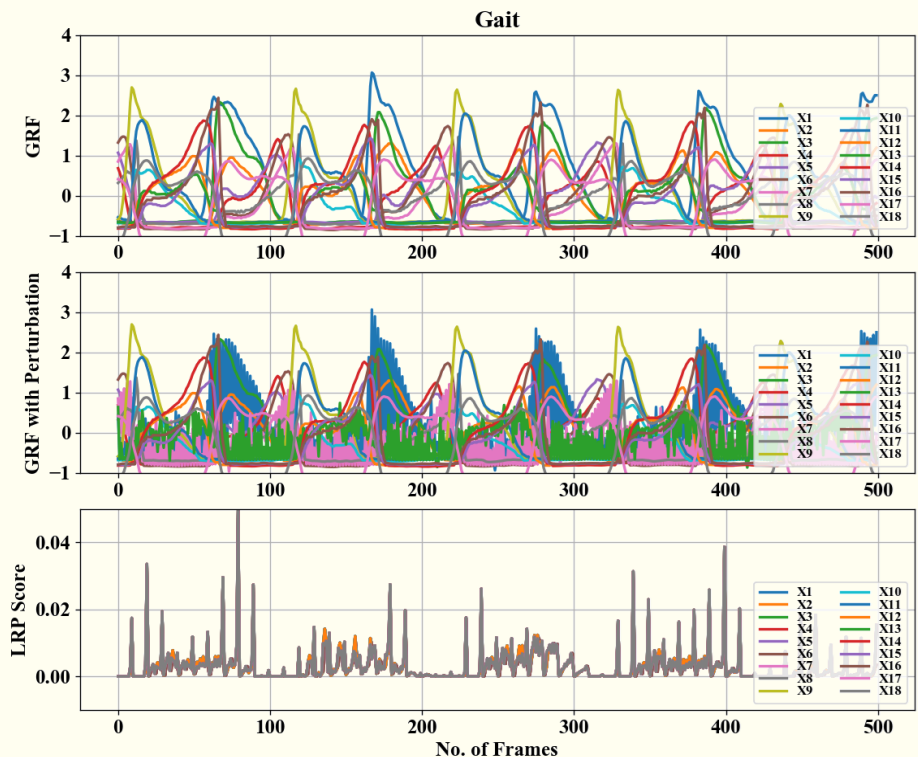

Figure 11. Perturbation with Gaussian noise based on LRP relevance scores, used for LRP and DCNN selection. Top plot: healthy gait processed GRF signals; Middle plot: GRF with Perturbation noise; Bottom plot: LRP relevance scores for the selected LRP Sequential Preset a Flat (LRP-SPF) and Parallel DCNN. $x_i = [x_1 \quad \cdots \quad x_{18}]$ represent the 18 signals after data standardization (see figure 6). The LRP plot is dominated by $x_{18}$ because all the signals are plotted on top of each other in the temporal domain. LRP relevance score are highly dependent on the temporal changes, whereas spatial variation does not affect the model prediction.

### 4.1.4 Gait Event Assignment Using LRP

Gait GRF data take the form of periodic sequences which are characterized as repetitive cycles for each foot. We note that the normal gait cycle is initiated by the heel strike of one foot, followed by other gait events described in figure 4, in strict order. Therefore, the LRP-generated heat map of the temporal variations in the GRF signal can reveal which events in the gait cycle are most relevant for the classifications. Consequently, gait event assignment is best performed on the data sequences in figure 6 after spatial averaging and standardization. A representative spatially averaged sensor signal sequence is shown in figure 4.10 (a) for a healthy subject. The highlighted gray area corresponds to one gait cycle, while the plotted signal is given by the Spatial Average (SA) metric, computed as:

$$SA[n] = \frac{1}{18}\sum_{i=1}^{18}(x_i[n] \tag{12}$$

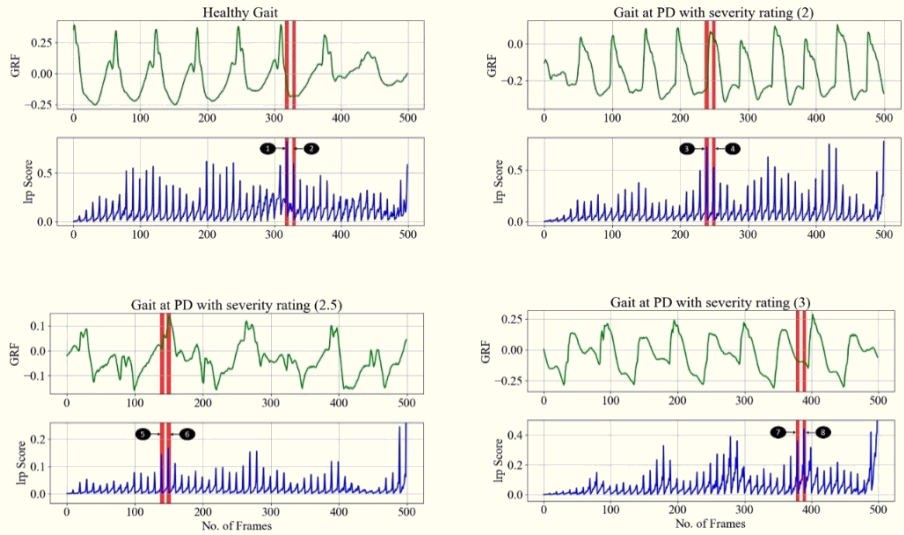

Figure 12. LRP method applied on randomly selected samples for the four PD severity ratings. SA of gait spatiotemporal signals: green; SA for LRP relevance scores over the same temporal period: blue. Vertical red bars with number labels display consistency with gait events listed below with capital letters as per figure. 4 and 10: 1, 3 and 6 - Heal strike and foot flattening (A); 2- Mid-stance and single support (C); 4- Loading response after the double support interval (B), 5 and 8 - Terminal swing and ready for the heel strike (G), 7- (F) Initial swing and Mid-swing or toe-of.

Here $x_i$ are the readings from individual sensors and $n$ enumerates the frames in each sample. Recall that each foot has 8 sensors attached (16 total) and the two sums one for each 8 sensors for each foot is available giving 18 signals in total. Figure 10 (b) shows the expanded gait cycle from figure 10(a) with the gait events color-coded and labelled as per figure 4.

The random data sample in figure 11 is used to illustrate the choice of the LRP method by the perturbation approach, as well as the LRP relevance scores obtained for the Parallel DCNN classifications. It shows the processed original GRF signals in the top plot; the middle plot shows the regions replaced with Gaussian noise, in view of the relevance scores shown in the bottom plot. While temporal data patterns yield classifications, the temporal maps of LRP scores highlight data intervals most significant for a given class. The plot of LRP scores consists of sharp peaks, well defined in the temporal domain, thus attributable to time-stamped gait events. Figure 12 displays the spatially averaged data signals for the four classes with their respective LRP score maps. The most prominent peaks are attributed to observable gait events, labelled in consistence with the gait cycle in figure 4. These are further discussed in section 5.

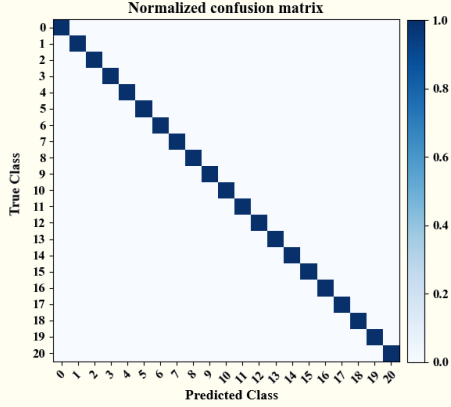

Figure 13. Gait signature classification confusion matrix for 21 subjects. The diagonal squares are the true positives, in this case 100% of classification and elsewhere are the false positives (0%)

### 4.2 Experiment (2) on Cognitive Load Gait Data

#### 4.2.1 Data Pre-Processing and Feature Learning and Classifications

In this experiment gait analysis is handled as a supervised learning process. Here, we propose a CNN model, based on the above extensive experimentation, as automatic feature extractor and classifier. The model shown in figure 5.a maps the gait spatiotemporal signal $\widehat{x_{n,s}}$ to an output label $y$ by learning an approximation function $y = f(\widehat{x_{n,s}})$.

Similar to experiment (1) An Adam is utilized to train and validate the model (for several experiments) using a batch size of 100 samples for each iteration; 200 epochs are found optimal to train the model. The training and validation sizes are set to be 70% and 10% respectively, where 20% is reserved for testing the model accuracy. The model is trained, validated, tested for several runs with data split using different random state parameters with different seeds. The mean performance and standard error are used to report the accuracy as follows:

$$SE = \frac{\sqrt{\frac{\sum(F1-\mu)^2}{q}}}{\sqrt{q}} \qquad (13)$$

A set of measured data as $x_{n,s} = [x_{n,1} \quad \cdots \quad x_{n,116}] \in \mathbb{R}^{n \times 116}$ is harvested from the iMAGiMAT system, where $n$ is the number of the data block (100 frames) and $s$ enumerates the POF sensors. A total number of 1050 samples are recorded for 21 subjects and placed in a 3D matrix of dimensions $1050 \times 100 \times 116$. The recorded amplitude of data varies due to the weight of each subject, therefore, data standardization is implemented as a pre-processing step, to ensure that the data is internally consistent, such that the estimated activations, weights, and biases update similarly, rather than at different rates, during the training process and testing stage. The standardization involves rescaling the distribution of values with a zero mean unity standard deviation, using equation (12):

The proposed model is trained, validated, and tested on m×n×s (m=number of samples, n=number of frames, S=number of POF sensors) spatiotemporal samples as K×100×116 for several runs using different *random state* parameters. m is chosen on the basis of experimental protocols and the mean performance and standard error are used to calculate the accuracy. Experiments are conducted to investigate the ability of the deep CNN to identify gait signature patterns by fusing 116 POF sensors in the model's deep layers, to extract gait patterns automatically in the following experiments.

Table 4. Models Classification Accuracy for Each Subject.

| Subject number | F1- score | Subject number | F1- score | Subject | F1- score |
|---|---|---|---|---|---|
| 0 | 95% | 7 | 87% | 14 | 100% |
| 1 | 65% | 8 | 90% | 15 | 75% |
| 2 | 93% | 9 | 90% | 16 | 80% |
| 3 | 90% | 10 | 77% | 17 | 100% |
| 4 | 87% | 11 | 91% | 18 | 100% |
| 5 | 91% | 12 | 90% | 19 | 80% |
| 6 | 73% | 13 | 100% | 20 | 69% |

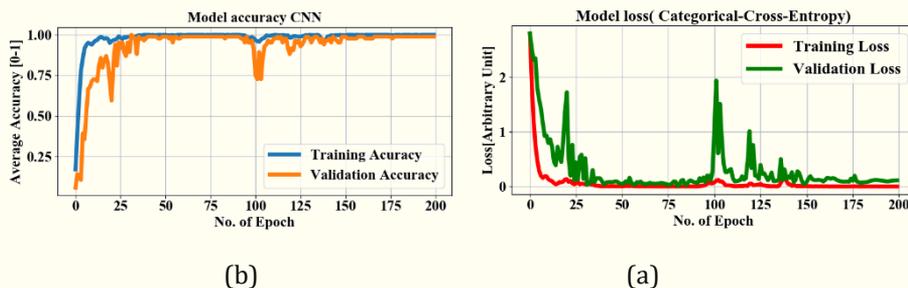

(b)        (a)

Figure 14. Model training and validation loss in (a) and model accuracy in (b) for 21 subjects in 5 classes

### 4.2.1.1 (21) Subject gait signature verification

To demonstrate the model's ability to verify the identity of a subject based on their gait signature, we assigned each subject's data a label numbered from 0 to 20, containing 50 samples of normal and cognitive load as explained in section 3.3.1. The model is trained, validated and tested on m=1050 samples with different *random state* parameters, and the mean performance and standard error are used to calculate the accuracy. The median classification confusion matrix is shown in figure 13, where the model achieved F1-score of 100% prediction and mean performance and standard error of 99.5±0.28 %. Figure 14 demonstrates the learning curve performance of the CNN over the iterations while training. We generate the training loss for each of the training sets and validation loss for each of the validation sets over the epochs. Figure 14 (a) shows the average training and validation losses.

Table 5. F1-score predictions for binary classification.

| Data Group for Classification | 1 testing subject | 2 testing subjects | 4 testing subjects | Test with all subjects |
|---|---|---|---|---|
| Class 0 vs class 1 | 100% | 85% | 81% | 79% |
| Class 0 vs class 2 | 95% | 87% | 58% | 69% |
| Class 0 vs class 3 | 60% | 68% | 63% | 79% |
| Class 0 vs class 4 | 100% | 85% | 74% | 81% |

The training loss starts from 3 and gradually reduces to 0. The validation loss generally follows the training loss, with a few spikes, and stablizes after 150 epochs. As expected, accuracy increases with decreasing loss, demonstrated in figure 14 (b). The average training and validation accuracy stablizes after 150 epochs after a few spikes.

For additional testing of the model performance in real-life scenarios, we evaluate the model on imposter and client classification. The client's data are used for the model training and validation and only 20% of that data is used for testing, while the imposters data are only used at the testing stage. The model at the testing stage predicts the client's gait sample identity with an F1 score of 100% and unable to predict the imposter which returns 0% F1 score. This is achieved by taking 17 subjects as clients (m=850 samples), and 4 subjects as imposters (m=200 samples). Clients were split 70%-10%-20% for training validation and testing respectively. In the testing stage, the model was able to correctly distinguish imposters and clients in 100% of cases.

### 4.2.1.2   Gender Classification

To demonstrate the model ability to recognize gait signatures, we perform a two-class classification based on the gender of the subject using the normal gait and cognitive load samples. The model is trained and validated with 6 subjects (m=300 samples) including 3 males and 3 females. Model testing is by predicting the gait class of two new subjects (m=100 samples, never seen by the model), the selection of males and females are done randomly. In this experiment the deep CNN prediction achieved F1-score of 95%, with 96% true positive prediction for the male samples, and 94% true positive prediction for the female samples.

### 4.2.1.3   (21) Subjects Cognitive Load Classification

The aim of this experiment is to show that in healthy subjects the influence of cognitive load on gait varies from subject to subject and the normal gait can be predicted with higher true positive rates than predictions under cognitive load. Five types of gait signatures, normal and four cognitively demanding task patterns, are learned for 21 subjects. The performance observed for the 5 classes is shown in figure 15, as the median confusion matrix based on several runs with F1-score of 50% and mean performance and standard error of 48.25±1.03%.

The results show that normal gait is predicted by a true positive incidence of 92%, while there is notable confusion between the dual tasks performed by the 21 subjects. The different *random state* parameters return the same result, where the normal gait true positive prediction is higher than 90% and substantial confusion between the dual task cases. Figure 16 shows the CNN learning curves over the training iterations, where the training loss declines from 1.8 to 0 while the validation loss rises from 1.8 to around 4 for 200 epochs, resulting in low validation accuracy as per figure 16 (b) to evidence severe overfitting.

Table 6. F1-Score Predictions for Comparison of CNN with Classical Classifiers.

| Classifier | Experiment 1 | Experiment 3 |
|---|---|---|
| SGD | 77% | 42%, N=47% |
| KNN | 87% | 51%, N=81% |
| GPC | 5% | 22%, N=0% |
| CNN | 100% | 50%, N=92% |

N: True positive prediction of normal gait.

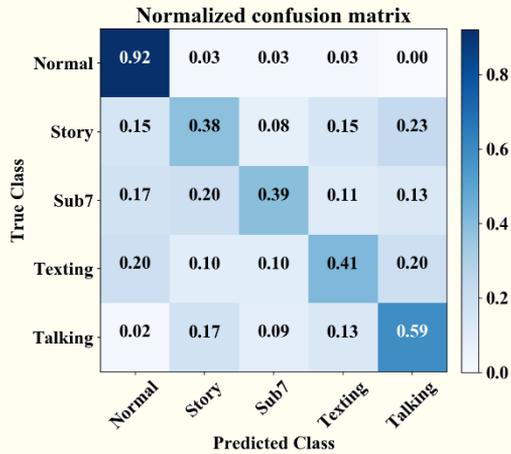

Figure 15. Confusion matrix for classification under cognitive load: 21 subjects, 5 classes.

### 4.2.1.4 Single Subjects Cognitive Load Classification

In this experiment gait patterns are investigated within each subject, to show that each subject gait under cognitive load can be learned and predicted. This is achieved by training, validating, and testing the CNN to classify each subject gait pattern using the normal gait and cognitive load. Each subject data is split using *random state* to cover all the 5 classes for testing with m=50 samples. The model evaluation using the F1- score is detailed for each subject in table 4. Gait data is predicted with more than 85% F1-score for 16 subjects, and for 6 subjects F1-scores are between 65% and 77%.

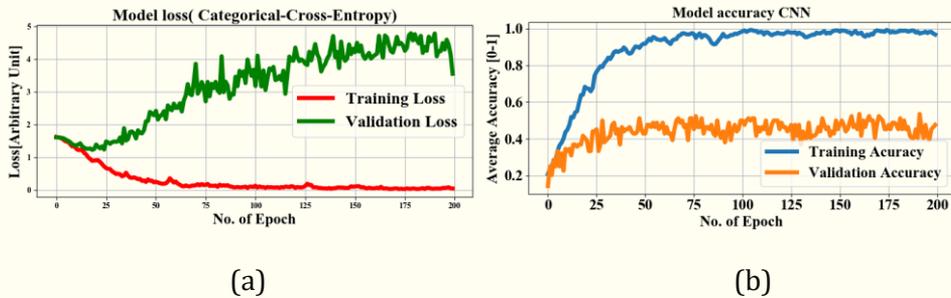

(a)            (b)

Figure 16. Model loss (a) and training and validation accuracy (b) under cognitive load, 21 subjects and 5 classes.

### 4.2.1.5 Binary Classification Under Cognitive Load

To study patterns for each of the 4 dual tasks (M2-M5) representing variants of cognitive load, we organize the data into four groups so that binary classification performance to distinguish between gait under normal (class 0) and cognitive load (one of classes 1, 2, 3 or 4, depending on the particular data group) conditions can be studied separately for each dual task. The CNN is trained 16 times, implementing 4 runs with each of the 4 data groups. The F1-scores for each run are shown in table 5. The first run in each data group is based on training and validating the CNN on 20 subjects and test the model on 1 subject, to see if we can predict gait of one person from 20 people. In the second run, the numbers are 19 and 2, respectively; in the third – 17 and 4 respectively. The last run is based on splitting the data into 70% for training, 10% for validation and 20% for testing, using m=420 samples with *random state* of 200 seed parameters (since the accuracy doesn't change with the *random*

*state seed*). As shown in table 5, the highest classification performance is achieved in the first runs (except for the group containing class 3). This is used essentially in the implementation of LRP to analyse the gait classes for that subject in the first run, as reported further comparison with Statistical Classifiers.

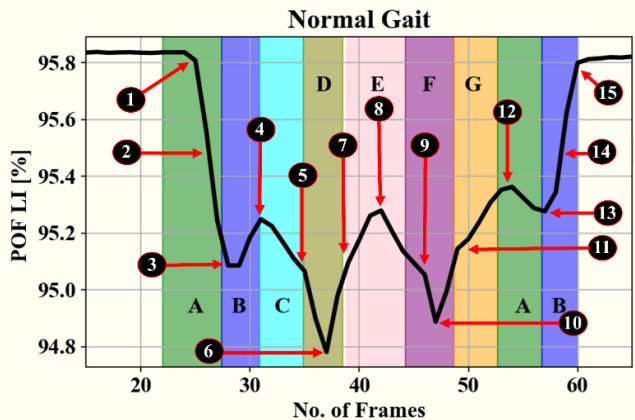

Figure 17. Representative gait cycle spatial average of a spatiotemporal signals (see equation 13). Gait events recorded by the sensors in a typical full gait cycle of two steps (figure 4: A,B,C,D,E,F,G): 1- Heel strike, 2- Foot-flattening, 3- Single support, 4- opposite Heel strike, 5- opposite Foot-flattening, 6- Double support, 7-Toe-off, 8- Foot swing, 9- Heel strike, 10- Double support, 11- Toe-off, 12- Foot swing, 13- opposite Heel strike, 14- Single support, 15-Toe-off.

### 4.2.2 Comparison With Statistical Classifiers

Gait signature recognition achieved high accuracy compared to Cognitive Load Classification; therefore, the validity of these achieved classifications is verified with statistical classifier. Here we compare the classification results achieved by CNN in experiment 1 and 3 with statistical classifier algorithms, such as Stochastic Gradient Descent (SGD) [198], K-Nearest Neighbors (KNN) [199], and Gaussian Process Classifier (GPC) [200]. To change the format of the statistical classifiers input, the data are flattened to length 11600=100×116, with and samples of m =1050 for experiment 1 and 3. The classification F1-scores on experiment 1 and 3 are shown in table 6. GPC fails in the true positive prediction of normal gait, while KNN achieves the best classification results. However, CNN outperforms the statistical classifiers for both gait signature recognition and normal gait prediction.

### 4.2.3 LRP Analysis of Gait Spatiotemporal Classifications

The focus of this section is to identify the features picked up by the model to classify gait under cognitive load. To obtain accurate LRP relevance scores $Ri$, the model true positive prediction should be high. Therefore, the gait class with a high positive rate is considered for LPR analysis. The learned CNN model parameters in experiment 1, 3 and 5 were frozen for LRP analysis. Experiment 4 is to check if there is a variation in gait within a subject; therefore, it is not considered for LRP analysis. LRP Sequential Preset a Flat (LRP-SPF) based on the LRP was utilized for this work, as it has shown sensitivity to gait inconsistency using perturbation in PD case.

The iMAGiMAT system captures a sequence of periodic events as distinct, but similar cycles for each foot. This spatiotemporal sequence is generated by the change of light transmission intensity in the POF sensors: $x_i = [x_1 \quad \cdots \quad x_{116}] \in \mathbb{R}^{n \times 116}$. However, a typical interpretation of the gait cycle, based on visual observation, is derived much less from the spatial component than the temporal one.

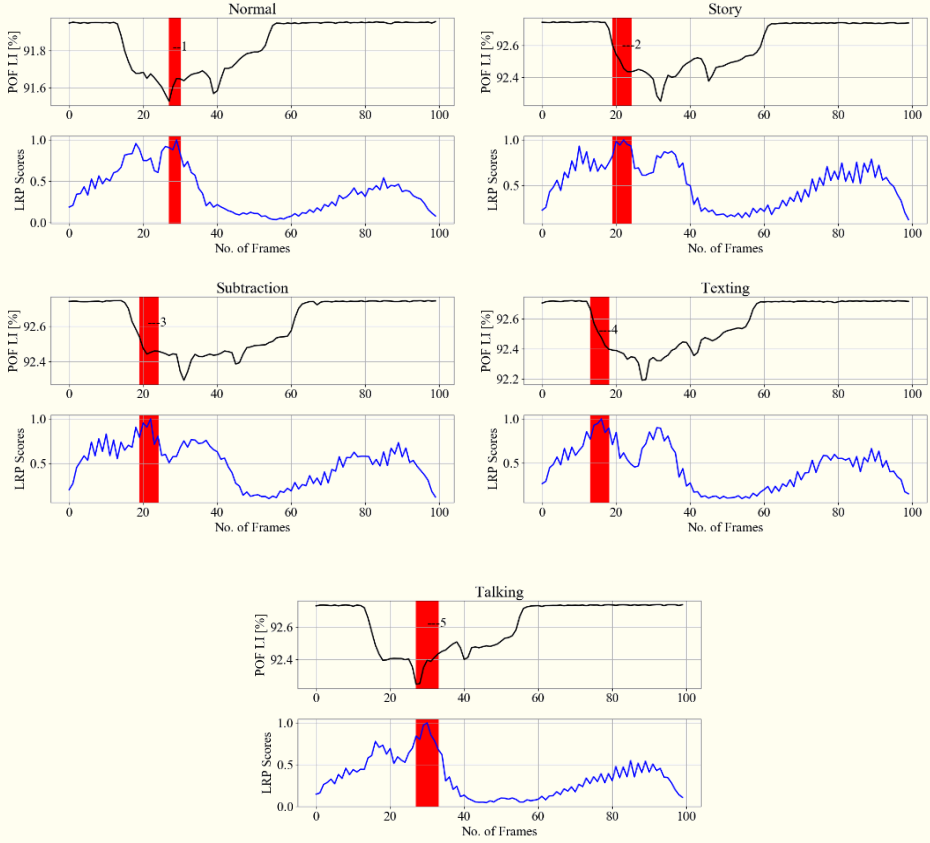

Figure 18. LRP methods applied on a single subject from experiment 1 testing data (each column is one pair), to identify gait events relevant for the CNN prediction to classify the cognitive load impact on gait. SA of gait spatiotemporal signals: black; SA for LRP relevance signals over gait temporal period: blue; POF LI (Plastic Optical Fiber Light Intensity). Vertical red bars with numbers display correspondence to gait events as per figure 17: 1,5- Loading response or Foot flat and Double support, 2,3,4 - Loading response or Foot flat and Single support.

Thus, to progress towards interpreting the CNN classifications in terms of observable gait events, we average over the spatial domain according to:

$$SA[n] = \frac{1}{s}\sum_{s=1}^{116}(x_{n,s}) \qquad (14)$$

Here $x_{n,s}$ are the readings from individual sensors $s$ at a specific frame $n$ within each sample and $SA$ is the frame $n$ spatial average calculated as the arithmetic mean over all sensors. Figure 17 displays a typical $SA$ of the spatiotemporal gait signal, labelling the main gait events over a two-step gait cycle.

Figures 18 and 19 display randomly chosen samples of single subjects, returning 100% true positives prediction for gait signature verification in experiment 1; figure 20 displays randomly selected samples of normal gait classified with 100% true positives in experiment 3; figure 21 shows predicted gait samples in experiment 5 for a subject never seen by the model when the training set is 20 subjects.

The top panels in figures 18, 19, 20 and 21 display calculated $SA$ aligned against the relevance "heat map", generated from the calculated LRP scores and displayed in the bottom panels (to be discussed further in section 5). The SA temporal sequences have different values on the *y* axis due to the nature of the captured gait signal, which is influenced by the individual anthropometry of subjects.

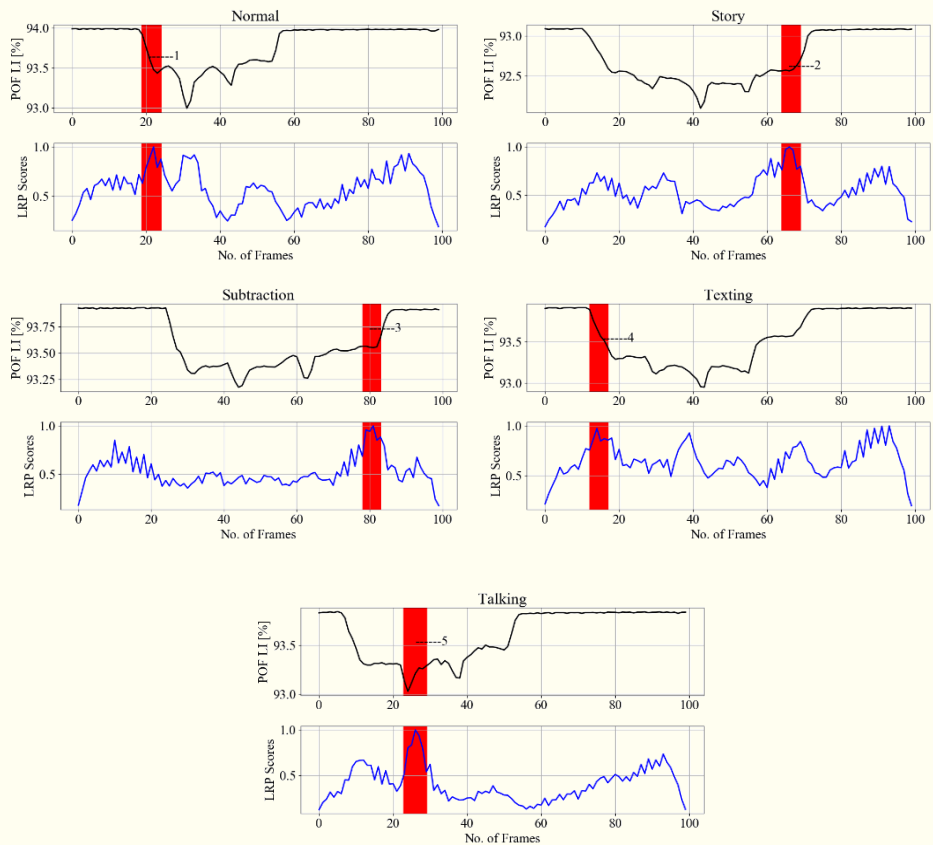

Figure 19. Consistent with identifying gait events relevant for the CNN prediction, random subjects from experiment 1 gait events are: 1,4- Loading response or Foot flat and Single support, 2,3- Foot swing and opposite Heel strike, 5- Loading response or Foot flat.

Table 7. PD classification results on PhysioNet three datasets.
SVM: Support Vector Machine; 1D-LBP+MLP: Shifted 1D-Local Binary Patterns + Multi-Layer

| Reference | Methods | Accuracy [%] |
|---|---|---|
| E. Abdulhay et al. [63] | SVM | 92.7 |
| Y. N. Jane [64] | Q-BTDNN | 91.5 |
| Ertuğrul et al. [65] | 1D-LBP+MLP | 88.89 |
| Medeiros et al. [66] | PCA | 81.00 |
| Wu et al. [67] | SVM | 84.48 |
| This work | Parallel 2D-DCNN | **95.5±0.28** |

Perceptron; PCA: Principal Component Analysis; Q-BTDNN: Q-back propagated time delay ANN.

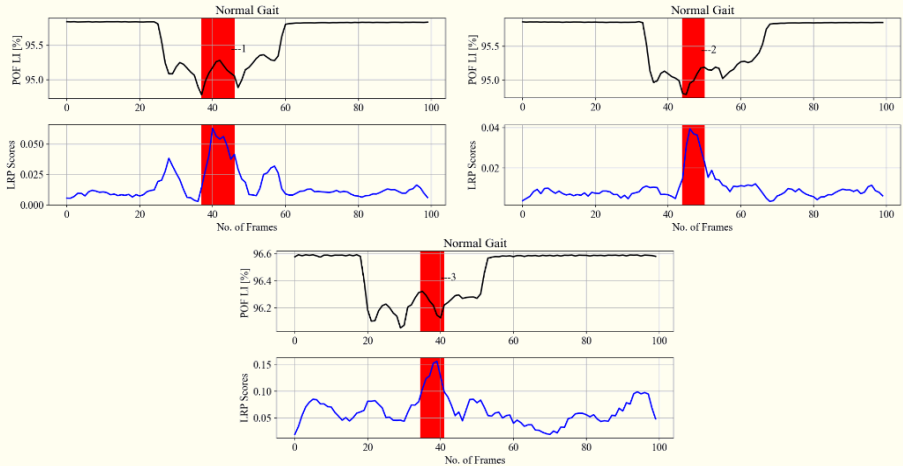

Figure 20. LRP methods applied on normal gait samples (from different subjects) from experiment 3 testing data, to identify gait events relevant for the CNN prediction to classify the cognitive load impact on gait. Gait events are: 1,2,3- loading response or Foot flat and Double support.

## 5. Discussion

The study presented delves into the promising realm of explainable artificial intelligence (AI) and deep learning methods for predicting gait deterioration. The focus is on identifying the impact of cognitive load and Parkinson's disease (PD) on gait patterns, and this is achieved by analyzing spatiotemporal data obtained from sensors placed under the feet. To carry out this investigation, Convolutional Neural Networks (CNNs) were utilized. These powerful neural networks can effectively learn from complex spatiotemporal data and produce highly accurate predictions. In addition, the CNNs were perturbed to provide insights into the features within the spatiotemporal gait ground reaction force (GRF) signals that are most relevant to the models' predictions. The results of this study are presented in detail in the following sections, with each data classification and perturbation analyzed and discussed in depth.

### 5.1 PD Data

The spatiotemporal signal in figure 6 implies that gait has normal events. Abnormal gait, otherwise difficult to detect visually, can be detected by machine learning, in alignment with the knowledge of the ground truth labels. However, the magnitude of GRF in newton shows a decrease attributable to the severity of PD. The main objective of this work is to find the best deep learning model for PD severity rating and relate the model predictions to the gait cycle events shown in figure 4 and figure 10.

Research towards machine learning classifications from PD data specifically PhysioNet data is based on the use of manual feature extraction methods with the classical machine learning methods as shown in table 7. The best classification results from manual extraction are reported in [63] using SVM classifier (92.7%). Our previous work on PD severity classification [58], reported that the 2DCNN outperformed the 1DCNN, SVM, decision tree algorithm, logistic regression algorithms, multi-layer protection and LSTM. In this article, we explore three architectures for automatic extraction and LRP analysis. The proposed DCNNs identified PD, as well as rated the severity of the deviation from healthy gait, achieving better classification performance with F1-score of 98% for each dataset and for the datasets combined with different random state (see table 3). The best classification accuracy is achieved with the parallel 2D-DCNNs, with mean performance and standard errors of 95.5% and 0.28%, respectively. Additionally, the parallel 2D-DCNNs exhibit robustness at perturbation with Gaussian noise as shown in figure 9. This suggests that the model is adequate for detecting gait deterioration from the spatiotemporal GRF signal. As an additional substantial enhancement, our LRP approach allows classification results to be related to visual observations similar to those established in medical practice to diagnose PD.

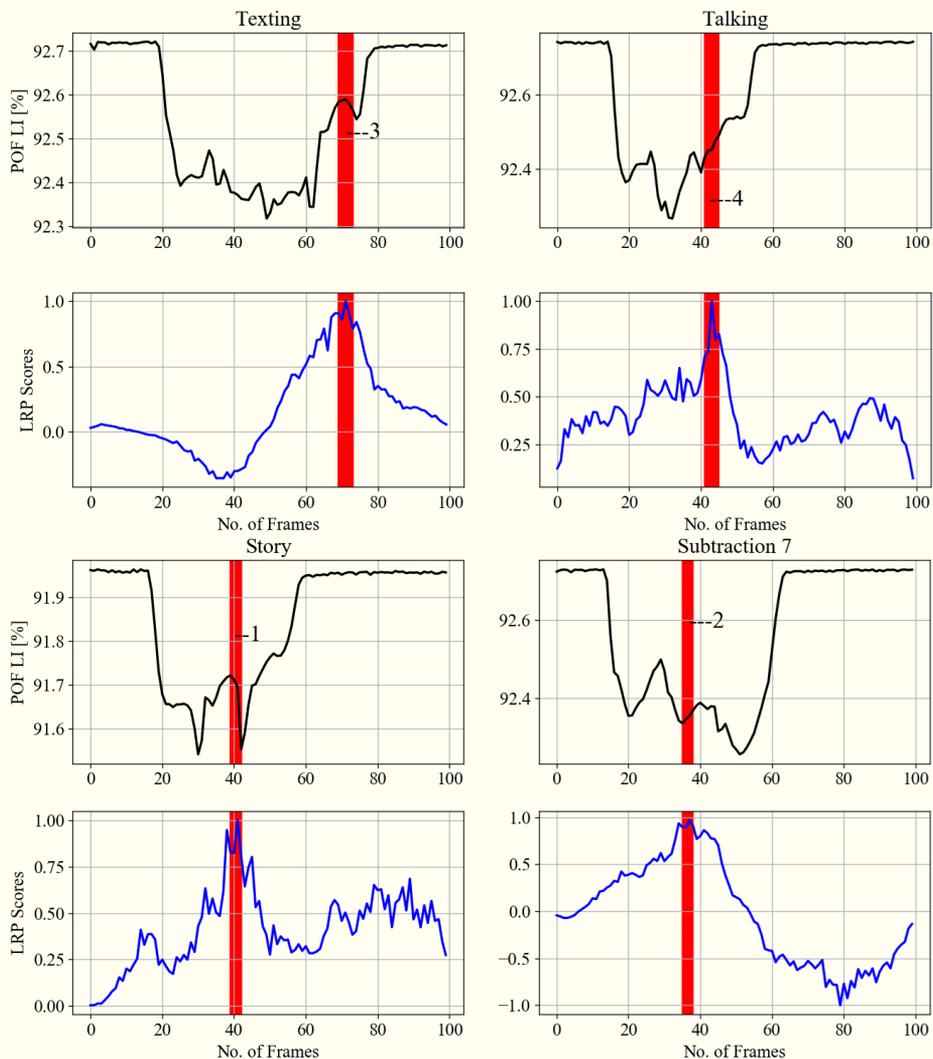

Figure 21. LRP methods applied on a single subject from experiment 5 testing data (each column is one pair), to identify gait events relevant for the CNN prediction to classify the cognitive load impact on gait. Gait events are: 1- Heel strike, 2- Toe-off, 3- between Foot swing and opposite Heel strike, 4- between Double support and Toe-off.

The DCNN classifies the raw spatiotemporal signals as healthy or within three severity ratings as shown in the confusion matrix (figure 7). The best LRP method is selected by applying a perturbation technique, which detects the highest sensitivity to removal of information from the input data sequence (figure 8). The selected LRP-SPF was found to be superior to well-known methods such as deconvolution and guided backpropagation.

Among the DCNN architectures (figure 5), the Parallel DCNNs model shows the steepest decrease in the perturbation procedure. Therefore, that model is learnt and used to generate the heat map or relevance randomly selected samples (figure 12), without distinction between left and right foot signal is for two feet. The gait cycle events identified as key at each level of PD severity are listed below:

1) **PD Severity Level 0 (Healthy Gait):** 1- Heel strike and foot flattening (A), 2- Mid-stance and single support (C). This indicates that the healthy person's ability to maintain balance is stronger than the PD patients', with strong balance suggesting that the forces are applied rhythmically to achieve the lower limbs' synchronized movement with stable posture.
2) **PD Severity Level 2:** 3- Heel strike (A), 4- Loading response after the double support interval (B). The heatmap shows that the subjects affected with PD level 2 have a weaker heal strike followed by a weaker balance in double support, where this feature is marked by the model by 96% f1-score.
3) **PD Severity Level 2.5:** 5- Terminal swing (G), 6- Heel strike (A). This shows that the subject has weaker foot landing or flat foot landing after the balance is compromised by the single support.
4) **PD Severity Level 3:** 7- Initial swing and mid-swing or toe-off (F), 8- Terminal swing and ready for the heel strike (G).

Here the balance is compromised by weak GRF resulting from unstable body posture and implies high risk of falling. This conclusion is based on linking the stages of PD in [68] (description of how the stage of PD is affecting the body poster during gait using visual observation) to the events that are highlighted by the model for a certain PD severity.

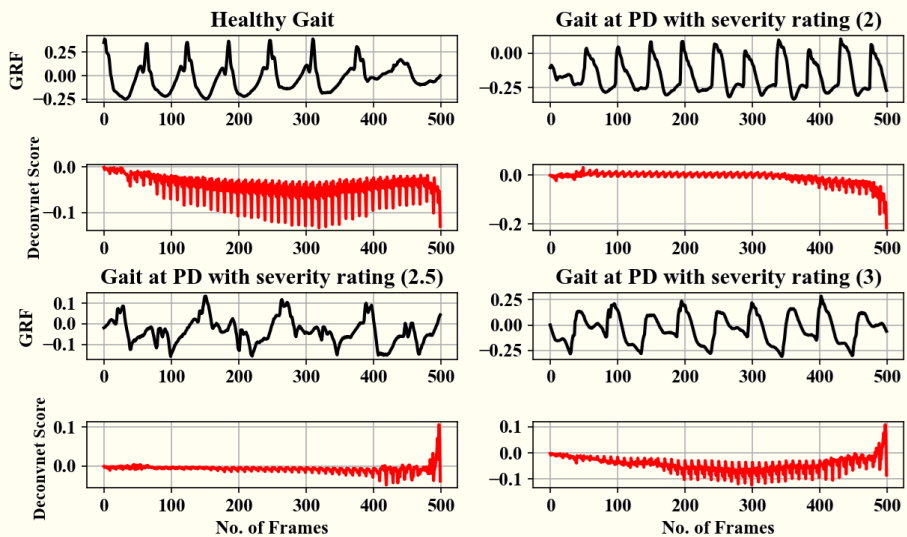

Figure 22. A deconvolution decomposition method applied to explain the parallel DCNN model prediction for PD severity ratings. SA of gait spatiotemporal signals is in black plot and model decomposition using deconvolution for each class in red plot. The deconvolution plot spikes every 10 frames.

The above markers for classification align with the observations in literature that PD-induced gait GRF deterioration affects the body balance and posture. The latter are with the closest relevance to gait events identified by the heat maps in figure 8 as the highest LRP scores, while the other gait events are less significant to the classifications. It is worth mentioning that these markers are identical by 95.5% in 1281 samples, such that the removal of these regions in the 95.5% of samples resulted in strong decay in the model prediction. The interpretation given above is in very good agreement with the description of the Hoehn and Yahr Scale staging criteria, as follows: "Stage 0 - No signs of disease, Stage 2 - Symptoms on both sides but no impairment of balance, Stage 2.5 - Mild symptoms on both sides, with recovery when the 'pull' test is given (the doctor stands behind the person and asks them to maintain their balance when physically pulled backwards), Stage 3 - Balance impairment, mild to moderate disease, physically independent" [68]. However, the staging criteria do not refer to the gait events adversely influencing the body's postural balance, due to the advancement of disease.

The analysis of LRP score in figure 11 and 12 reveals a consistent spike in the LRP plot for every 10 frames, thus further investigation has been carried out by plotting the model decomposition using deconvolution, Deep Taylor and LRP-SPF for a single stream DCNN model. As shown in figures 22 and 23, the spikes are also consistent in all the plots. Further, the sensors used to record PD and healthy gait in [45],[46],[47] are pressure-sensitive sensors to measure the forces underneath the foot as a function of time at a rate of 100 Hz.

These spikes can be an artefact generated from the data processing, either in the forward pass (classification) or the backward pass (LRP decomposition) due to the pooling lyears. However, the data is considered reliable, based on its noise resilience as demonstrated in our perturbation analysis.

## 5.2 iMAGiMAT Data

### 5.2.1 Classification of Gait Signatures under Cognitive Load

The present study investigates the importance of cognitive load influence for gait inconsistency. We present a comparison of classification performance between 5 types of gait: normal and under cognitive load in 4 different tasks. Deep CNNs not only outperform, unsurprisingly, the classical classifier methods but also achieve an F1-score of 100% (see figure 13 and table 4) for gait signature verification in experiment 1 with 21 healthy adult's data, and 100% prediction of 4 imposters and 17 clients. The learning curve in figure 14 demonstrates the good match of the CNN methodology for gait verification tasks. The network parameters are updated via backpropagation to map gait during training to 21 classes are correctly optimized at the validation stage, which is important for the testing stage to make prediction for gait verification.

Experiment 2 is in essence an extra validation of the adequacy of the spatiotemporal sampling of GRF by the 116 sensors and their fusion, as well as that the classification performance of the trained models. An F1-score of 95% is achieved for test data from an unseen male as well as an unseen female. Although experiment 2 has the character of a sanity check, the results lend support to the value of floor sensor gait data as a biometric.

Experiment 3 is conducted to study the possibility to classify cognitive load on healthy subjects. It has shown that normal gait is classified with a higher true positive rate compared to any of the classes of gait under cognitive load. This experiment also indicates that the achieved true positive rates in predicting normal gait are higher for the CNN model compared to the classical classifiers (see figure 15 and table 6). The learning curve in figure 16 indicates overfitting [69], to imply that the gait patterns under cognitive load diverge among the 21 subjects. Samples obtained under cognitive load samples are hard to fit due to the inconsistency of gait pattern changes among the subjects.

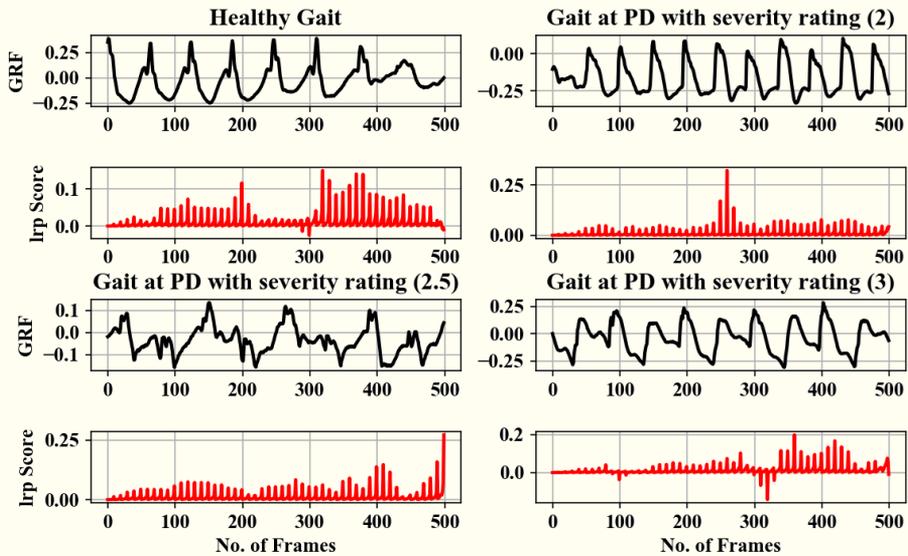

Figure 23. A Deep Taylor decomposition method applied to explain the parallel DCNN model prediction for PD severity ratings. Similar to figure 4.11 and 4.12 the Deep Taylor decomposition plot spikes every 10 frames.

The results from the first three experiments suggest that while the dual-task data obviously contributes to the high F1-scores in experiments 1 and 2, it results in substantially degraded true positive rates in experiment 3. However, experiment 4 shows that when classifications are within a single subject the performance is notably better: for 16 subjects (out of 21) the gait under cognitive load the F1 score ranges between 80 to 100%, with the remaining 5 subjects the range being between 69% to 77%.

These observations can be discussed in the light of humans having a natural gait pattern evolved over millions of years; however, changes in gait when experiencing cognitive load at any particular instance are specific to the individual, expressing their response to the impaired ability to process cognitive information. In experiment 5, we use binary classifications (see table 6.2) to distinguish normal gait from gait under the 4 variants of cognitive load. The best classification results are obtained when the model learns normal or dual-task gait features for a single subject. This implies that although learned gait features under cognitive load may not be readily portable across subjects, they are consistent for each individual and can contribute substantially for correct subject classifications; however, the accuracy drops if more subjects are involved.

### 5.2.2 Interpretation of Classifications

Figures 18, 19, 20 and 21 provide the link between the LRP relevance scores ("heat map") and the time sequence of the calculated *SA* signal in a single gait cycle window. The LRP score maxima are suitable pointers to the parts of the gait cycle which are most relevant for the classifications. For accurate heat maps of a specific gait class the model's true positive prediction in the confusion matrix must be close to 100% for most of the testing samples, which points to the results from experiment 1 (figures 18 and 19), experiment 3 for normal gait heat maps - in figure 20 and experiment 5 for a single subject predicted gait under the 4 variants of cognitive load - in figure 21. Focusing just on one complete gait period (two steps) is justified by the fact that on multiple repetitive occasions each subject will initiate a gait cycle (see full description

of the gait cycle figure 4) by performing a heel strike, strictly followed by other gait events described in figure 17 and ending in a toe off.

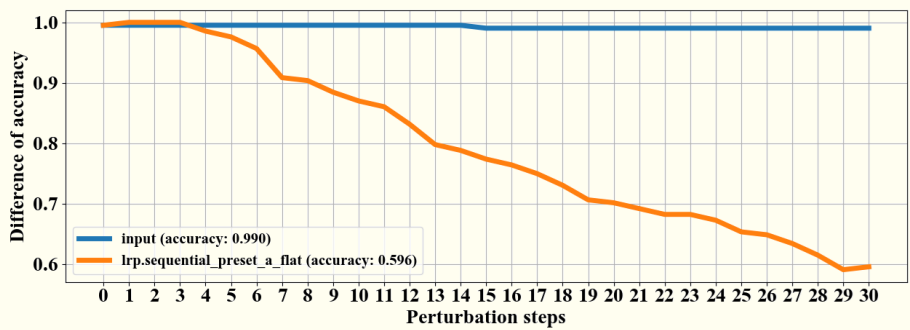

(a)

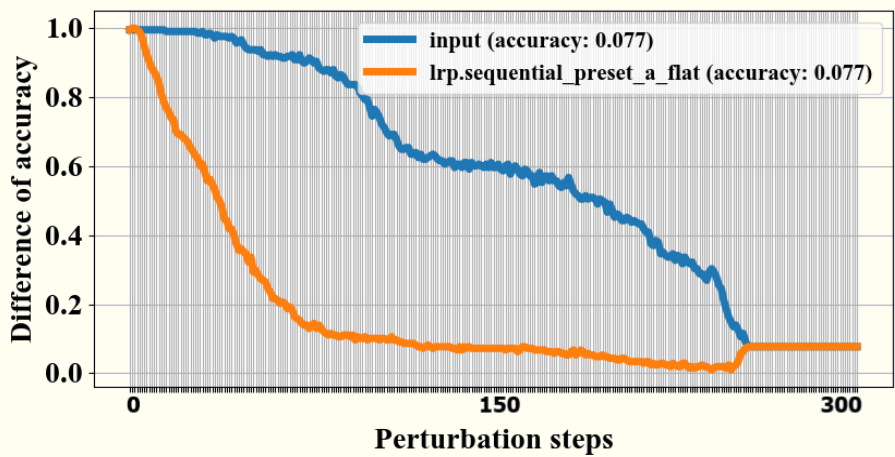

(b)

Figure 24. LRP heatmaps validation by perturbation technique for experiment 1. Information with the highest relevance scores is progressively removed and the test samples are re-predicted. Steeper initial decrease indicates better identification of gait events with most weight in the classifications. a) shows the model predictions in 30 steps based on removing relevance scores using LRP Sequential Preset a Flat (LRP-SPF) and random removal of information. b) shows the model performance after 300 steps of information removal.

Figure 18 indicates that loading response has high relevance to assigning a gait signature to one out of the 21 subjects gait samples, notably even under cognitive load, as indicated by with gait events numbered from 1 to 5. Figure 19 shows another subject randomly selected out of the 21. The indicated gait events are loading response for normal gait (1), gait while texting (4) and gait while talking (5); Foot swing and opposite Heel strike for gait while listening to a story (2) and while performing subtraction (3). The indication of events numbered 1, 2, 3 on figure 6.9 implies that normal gait identified by loading response or Foot flat and double support for 21 subjects. This gait event is marked by the model by 92% true positive (see figure 6.4) to distinguish normal gait from 4 cognitive load classes. Figure 20 shows cognitive load gait samples for one subject as per experiment 5 (the one subject never seen by the model) summarized as follows:

i. Gait while listening to story: Heel strike is significant for distinguishing listing to story from normal walking.
ii. Gait while performing serial 7 subtraction: Toe-off is significant for distinguishing 7 subtraction from normal walking.

iii. Gait while texting in smart phone: the transition from foot swing to opposite Heel strike is significant for distinguishing texting from normal walking.
iv. Gait while talking: the transition from double support to Toe-off is important to distinguishing talking from normal walking.

Overall, the LRP analysis indicates that subjects' normal gait is characterized by loading response, while the other cognitive load gait classes is classified by landing or lifting the feet on/from the surface of the iMAGiMAT system. For subject verification there are many second relevant scores are used to predict the identity of the subject based on gait signature.

Figure 24 shows the assessment of the validity of the LRP heatmaps for subjects' identification using cognitive load. Here we apply the removal of region based on both LRP Sequential Preset a Flat (LRP-SPF) MoRF and random region removal and re-predicting gait class. As shown in figure 24 (a) the model prediction strongly decays using the LRP for the removal of information compared to the removal of random information. Figure 24 (b) shows the model performance over 300 steps. It can be seen that the model reaches lowest performance accuracy where the gait classes have to take a random prediction. Furthermore, it can be inferred from figure 24 that the model is effective in finding the most relevant region to identify subjects and the LRP is consistent over the test samples.

## 6. Conclusion

To conclude, this research work highlights the effectiveness of deep learning models in accurately classifying gait deterioration in Parkinson's disease patients. The models surpass previous methods that rely on manual feature extraction and are also capable of withstanding perturbation noise, making them highly robust. The LRP analysis confirms that body balance is a critical aspect for diagnosing PD [68], with higher levels of the disease affecting a patient's ability to walk without the risk of falling. The identification of relevant gait cycle events can also aid clinical practitioners in diagnostics, either visually or through quantitative parameters derived from observations. The methodology proposed in this article has the potential to contribute to developing a strategy for personalized longitudinal monitoring of the progression of PD severity. Additionally, the study shows that floor sensors can be used to capture changes in an individual's unique gait signature due to cognitive load, providing potential for biometric and security applications. In healthcare, gait data from floor sensors can contribute to the detection of Parkinson's disease onset and fall risks. The future direction of this research may involve the inobtrusive sampling of subjects' gait under routine conditions over intervals spanning periods of physical and mental changes due to aging, which could contribute to earlier detection of disease onset. Overall, this study demonstrates the vast potential of deep learning models and Explainable AI in the field of gait analysis, which could significantly improve clinical practice and patient outcomes.